\documentclass[aps,prd,twocolumn,showpacs,superscriptaddress,groupedaddress]{revtex4-1}  % for review and submission
\usepackage{graphicx, color}  % needed for figures
\usepackage{dcolumn}   % needed for some tables
\usepackage{bm}        % for math
\usepackage{amssymb}   % for math
\usepackage{amsmath}   % for math
\usepackage{multirow}
\usepackage{enumerate}
\usepackage{comment}
\usepackage{tabularx}
\usepackage{booktabs}
\usepackage{placeins}
\newcommand\m[1]{\mathrm{#1}}
\newcommand{\no}{\mathrm}

\begin{document}

% The following information is for internal review, please remove them for submission
%\widetext
%\leftline{Version 01 as of \today}
%\centerline{\em CONFIDENTIAL}

% the following line is for submission, including submission to the arXiv!!
%\hspace{5.2in} \mbox{Fermilab-Pub-04/xxx-E}

\title{Optical losses in a high-finesse 300 m filter cavity for broadband quantum noise reduction in future gravitational-wave detectors}

\author{Eleonora Capocasa$^{1,2}$} 
\email[Corresponding author:]{eleonora.capocasa@nao.ac.jp}
\author{Yuefan Guo$^{3}$}
\author{Marc Eisenmann$^{4}$} 
\author{Yuhang Zhao$^{1, 11}$} 
\author{Akihiro Tomura$^{5}$} 
\author{Koji Arai$^{6}$} 
\author{Yoichi Aso$^{1}$} 
\author{Manuel Marchi\`o$^{7}$} 
\author{Laurent Pinard$^{7}$} 
\author{Pierre Prat$^{2}$} 
\author{Kentaro Somiya$^{8}$} 
\author{Roman Schnabel$^{9}$} 
\author{Matteo Tacca$^{10}$} 
\author{Ryutaro Takahashi$^{1}$} 
\author{Daisuke Tatsumi$^{1}$} 
\author{Matteo Leonardi$^{1}$} 
\author{Matteo Barsuglia$^{2}$} 
\author{Raffaele Flaminio$^{4,1}$}

\address{$^{1}$National Astronomical Observatory of Japan, 2-21-1 Osawa, Mitaka, Tokyo, 181-8588, Japan}
\address{$^{2}$Laboratoire Astroparticule et Cosmologie (APC), 10 rue Alice Domon et L\'eonie Duquet, 75013 Paris, France}
\address{$^{3}$Beijing Normal University No. 19, XinJieKouWai St., HaiDian District, Beijing 100875, P. R. China}
\address{$^{4}$Laboratoire d'Annecy-le-Vieux de Physique des Particules (LAPP), Université Savoie Mont Blanc, CNRS/IN2P3, F-74941 Annecy-le-Vieux, France}
\address{$^{5}$The University of Electro-Communications  1-5-1 Chofugaoka, Chofu, Tokyo, Japan}
\address{$^{6}$LIGO, California Institute of Technology, Pasadena, California 91125, USA}
\address{$^{7}$Laboratoire des Mat\'eriaux Avanc\'es, CNRS-IN2P3, Universit\'e de Lyon, Villeurbanne, France}
\address{$^{8}$Graduate School of Science and Technology, Tokyo Institute of Technology, 2-12-1 Oh-okayama, Meguro, Tokyo, 152-8551, Japan}
\address{$^{9}$Institut f\"{u}r Laserphysik und Zentrum fur Optische Quantentechnologien der Universitat Hamburg, Luruper Chaussee 149, 22761 Hamburg, German}
\address{$^{10}$Nikhef, Science Park, 1098 XG Amsterdam, Netherlands}
\address{$^{11}$Department of Astronomical Science, SOKENDAI, 2-21-1 Osawa, Mitaka, Tokyo, 181-8588, Japan}
\date{\today}

% ---------------------------------------------------------

\begin{abstract} 

Earth-based gravitational-wave detectors will be limited by quantum noise in a large part of their spectrum. The most promising technique to achieve a broadband reduction of such noise is the injection of a frequency dependent squeezed vacuum state from the output port of the detector, with the squeeze angle is rotated by the reflection off a Fabry-Perot filter cavity. One of the most important parameters limiting the squeezing performance is represented by the optical losses of the cavity. We report here the operation of a 300-m filter cavity prototype installed at the National Astronomical Observatory of Japan (NAOJ). The cavity is designed to obtain a rotation of the squeezing angle below 100 Hz.  After achieving the resonance of the cavity with a multi-wavelength system, the round trip losses have been measured to be between 50 ppm and 90 ppm. This result demonstrates that, with realistic assumption on the input squeeze factor and on the other optical losses, a quantum noise reduction of at least 4 dB in the frequency region dominated by radiation pressure can be achieved. 

\end{abstract}

%\ocis{000.0000, 999.9999.}
%\pacs{04.80.Nn, 95.55.Ym, 95.75.Kk}
\maketitle

\section{\label{sec:introduction}Introduction}

After the groundbreaking first gravitational wave observations\cite{PhysRevLett.116.061102,BNS}, Virgo and LIGO gravitational-wave detectors will undergo a series of sensitivity upgrades alternating with scientific data takings. In parallel the Japanese detector KAGRA is being commissioned and it will soon join the gravitational-wave detector network. 

The design sensitivity of Advanced Virgo \cite {0264-9381-32-2-024001} , Advanced LIGO \cite {0264-9381-32-7-074001} and KAGRA \cite {PhysRevD.88.043007} is expected to be limited in a large part of the spectrum by the quantum nature of the light, through its manifestations of the shot noise and radiation pressure noise. As pointed out by Caves in 1981, both shot noise and radiation pressure originate by the vacuum fluctuations entering the detector from its output port \cite{PhysRevD.23.1693}. The possibility to manipulate the quantum noise by injecting a broadband squeezed vacuum field \cite{SCHNABEL20171} with a frequency independent squeeze angle, from the detector's output port has been also proposed by Caves \cite{PhysRevD.23.1693}.
 
The production of broadband frequency independent squeezed vacuum is a mature technology and recently 15 dB of broadband squeezed vacuum field was observed \cite{sq15} and its effectiveness has been successfully tested first in a prototype \cite{40m} and then in full-scale detectors, such as the GEO600 \cite{sqgeo} and LIGO \cite{sqnat}.  
 As a first step, frequency-independent squeezed vacuum source will be injected in Advanced Virgo and Advanced LIGO, allowing to mitigate the shot noise at the expenses of an increase of the radiation pressure noise at low frequency. For the moment this does not represent an issue, since other noises are limiting the sensitivity in that frequency region. 

In order to obtain a broadband noise reduction, the injected squeezed vacuum has to undergo a frequency dependent rotation which counteracts the one induced by the optomechanical coupling inside the interferometer \cite{0295-5075-13-4-003}. In order to achieve an optimal noise reduction, the rotation has to take place at the frequency where the radiation pressure noise crosses the shot-noise, in a region around 40-70 Hz for Virgo, LIGO, and KAGRA. 

A technique to impress a frequency dependence on the squeezing ellipse consists in reflecting a frequency-independent squeezed vacuum off a detuned suspended Fabry-Perot cavity, known as \textit{filter cavity} \cite{KKK}. Such reflection induces a differential phase change on the upper and lower vacuum sidebands inside the cavity linewidth, resulting in a frequency-dependent rotation of the vacuum quadrature. The rotation frequency depends on the cavity storage time, proportional to the product of the finesse and the cavity length.  

The possibility to achieve high-levels of frequency dependent squeezing is mainly limited by the optical losses. In particular optical losses of the filter cavity, mainly due to mirror defects, are expected to degrade the squeezing at low frequency, in the region where the vacuum squeezed field experiences the rotation.

At present, squeezed vacuum rotation has been demonstrated in the region of MHz \cite {PhysRevA.71.013806} and kHz \cite {PhysRevLett.116.041102}. A 300 m filter cavity prototype is being developed at the National Astronomical Observatory of Japan (NAOJ), in the former TAMA interferometer infrastructure, with the goal of demonstrating frequency dependent squeezing, with an angle rotation below 100 Hz, in the region where the rotation is needed for Virgo, LIGO and KAGRA. The design of the cavity has been presented in a previous publication \cite{mio}.

We report here the operation of the cavity with a multi-wavelength control system, the measurement of the optical losses, their comparison with the expected values and their impact on the quantum noise reduction. The integration of the squeezed vacuum source is still on going. 

The article is organized in the following way. In the next section the description of the experimental setup is given. In section 3 the requirements for mirror surface quality and the mirrors surface characterization results are reported. In section 4 the measurement of the optical losses are presented. Finally, in section 5 the conclusions and next steps are presented.  

\section{\label{sec:setup} Experimental setup}     

\begin{figure}[h!]
	\centering{\includegraphics[width=0.5\textwidth]{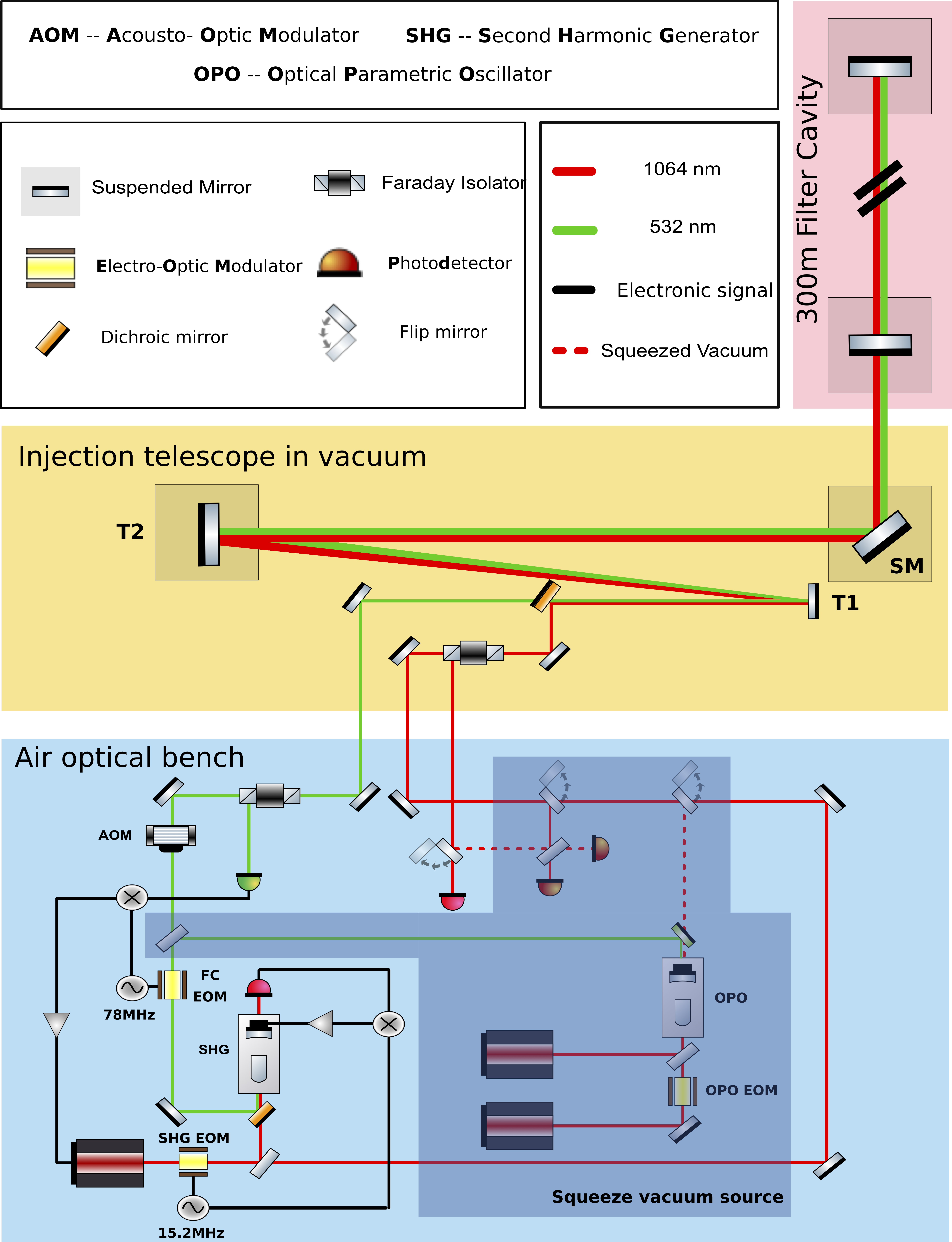}}
	\caption{Overall scheme of the experiment. The installation of the components in the grey shaded area on the optical table is currently ongoing. }
	\label{tamascheme}
\end{figure}

The experimental setup is composed of three parts: the source of a broadband squeezed vacuum field with frequency independent squeeze angle, the optics needed to inject the squeezed beam into the cavity, and the suspended cavity itself. The relevant parameters of the experiment are reported in Tab. \ref{recparfc} and an overall schematic of the setup is shown in Fig. \ref{tamascheme}

\begin{table}[!ht]
 \begin{center}    
 \begin{ruledtabular}
\begin{tabular}{llr}

Cavity parameter                                                                                                           & Design                       & Real value \\
\hline
Length                                                                                                                   & $300$ m                      & $300$ m \\
Mirror diameter                                                                                                      & $10 $ cm                      & $ 10 $ cm  \\
Input mirror radius of curvature                                                                             & 415 m                          & 438 m\\
End mirror radius of curvature                                                                               & 415 m                          & 445 m\\
Input mirror transmissivity  (1064 nm)                                                                   & $0.14\% $                     & $0.136 \%$   \\
End mirror transmissivity    (1064 nm)                                                                   & $<5$ ppm                     & 3.9 ppm  \\
Finesse   (1064 nm)                                                                                               &   $4290$                       &   $4425$\\
Input mirror transmissivity  (532 nm)                                                                     &   $1.4\% $                    & $0.7 \% $\\
End mirror transmissivity    (532 nm)                                                                     &   $1.4\%$                     &$ 2.9 \%$ \\
Finesse   (532 nm)                                                                                                 &   $445$                        & $172$  \\
Beam diameter at waist                                                                                         & $1.65$ cm                 & $1.68$ cm  \\
Beam diameter at the mirrors                                                                                & $2.06$ cm                & $2.01$ cm  \\

\end{tabular}
 \end{ruledtabular}
\caption{Summary of the filter cavity parameters. In the first column are reported the design values while in the second we reported the actual values.
Mirrors transmissivity and radius of curvatures (RoC) have been measured at LMA. The finesse has been estimated from the mirror transmission and assuming round trip losses of 60 ppm (as predicted by the results of the mirror characterization).}
 \label{recparfc}
 \end{center}
\end{table}

\subsection{\label{subsec: FISVS} Squeezed vacuum source} 

The design of the squeezed vacuum source (or \textit{squeezer}) is based on the design and experience of the GEO600 squeezer \cite{coco}. The main laser, a 2-W 1064-nm Nd:YAG laser,  is used to pump the second harmonic generator (SHG) cavity, which doubles the laser frequency producing green light at 532 nm.

In our scheme, this green light is split in two beams that are used for two different purposes. A part of the beam is used to pump the optical parametric oscillation cavity for the squeezed vacuum production. The other part is injected into the filter cavity and used to lock the main laser frequency to the cavity length by means of a Pound-Drever-Hall detection scheme \cite{PDHBlack}.  

The beam used for the control of the filter cavity passes through an acousto-optic modulator (AOM), which induces a tunable frequency shift and it is used to control the detuning of the infrared (squeezed) beam with respect to the cavity resonance. The infrared beam transmitted by the first beam splitter (placed before the SHG) is used in part as local oscillator for the homodyne detection and in part as a temporary test beam to characterize the filter cavity.

At the time of writing, the SHG has been assembled and locked. A part of the produced green beam has been used to lock the filter cavity and the infrared light transmitted by the first beam splitter has been superposed to the green one and made also to resonate into the cavity by using an AOM. The rest of the components are not yet integrated.

\subsection{\label{subsec: INJ} Injection path} 

As shown in Fig. \ref{tamascheme}, the infrared beam and the green beam are separately injected into the vacuum system.
The infrared beam passes through an in-vacuum Faraday isolator that is used to extract the infrared light reflected by the cavity. In the final setup this will be replaced by the frequency dependent squeezed vacuum beam and will be sent to the homodyne detector. In the present setup, the reflected infrared beam is sensed by a photodiode, in order to perform the loss measurements described in this article and to monitor its detuning from the cavity resonance. After passing through the Faraday isolator,  the infrared beam is superposed with the green beam by means of a dichroic mirror.
The superposed beams are then are magnified by a factor 10 using an afocal reflective telescope. This consists of two spherical mirrors (T1, and T2 in Fig. \ref{tamascheme}) with the radii of curvature of $-0.6 m$ (convex) and 6 m (concave), respectively, with a mirror distance of 2.7 m. After the telescope, the beams are reflected toward the filter cavity by a flat steering mirror with a diameter of 15 cm (SM in Fig. \ref{tamascheme}). The last mirror of the telescope and the steering mirror are both suspended. By adjusting their alignment, it is possible to align the beams on the cavity axis.

\subsection{\label{subsec: SUS} Filter cavity suspensions and alignment} 
Four suspended mirrors are used in our setup: two are part of the injection system described above, and two are the mirrors composing the filter cavity. The suspension system consists of a double pendulum originally developed for the TAMA experiment. It is composed by a top stage to which four wires are attached and used to suspend an intermediate mass. A passive damping system consisting of a set of magnets placed around this intermediate mass is installed. The mirror, with a diameter of 10 cm, is suspended with two loop wires from the intermediate mass. Four magnets have been glued to the back of the mirrors, allowing to move them by coil actuators. The double pendulum is placed on a vibration isolation multilayer stack made of rubber and metal blocks \cite{stack}.\\ The mirror position is sensed using optical levers with lenses which decouples shifts and tilts motion \cite{phd_eleonora}. The output of such systems is used as error signal for a local control loop which keeps the mirror motion in the range of few $\mu$rad and allows to align the cavity. After centering the green beam on the end mirror by using the suspended steering mirrors, the cavity is aligned by adjusting the position of the input and end mirrors to maximize the transmitted green power. At that point the infrared beam is aligned by using two steering mirrors on the optical table, in order to coalign it to the cavity axis and maximize its transmitted power.

\subsection{\label{subsec: FC} Filter cavity control} 
The filter cavity is kept resonant with the green beam using a standard Pound-Drever-Hall scheme in reflection and acting on the main laser frequency. 
The correction signal to the laser piezo is provided by an analog servo, with a bandwidth of about 20 kHz. In order to make the infrared beam also resonant in the cavity, the relative frequency of the green and the infrared beam is adjusted by driving the AOM placed on the green path. The frequency shift necessary to achieve the simultaneous resonance has been observed to be stable on the time scale of the green lock duration, that is few hours. In order to achieve the rotation of the squeezing ellipse, this value will be adjusted to operate the cavity at the proper detuning \cite{PhysRevD.90.062006, mio}. 
The cavity characterization has been performed using a bright IR beam which will be replaced with squeezed vacuum in the future. The lock accuracy, measured as the rms of the Pound-Drever-Hall error signal, is $\sim$120 Hz for the green beam. The lock accuracy for the IR, measured extracting an analogous Pound-Drever-Hall signal, is of the order of few Hz. The difference between the two lock accuracies is explained by the fact that the higher finesse of the cavity for the IR beam is filtering more the laser noise at high frequency, above the pole of the cavity, where the rms is mostly accumulated.

\section{\label{sec:mirrors} Cavity Mirrors requirements and characterization}  
The degradation of the squeeze factor induced by several loss sources has been modeled, taking into account losses in the injection and readout paths, effect of the mismatching  (between the squeezer and the cavity, and between the squeezer and the local oscillator), phase noise and filter cavity losses \cite{PhysRevD.90.062006}. 
Filter cavity losses have been shown to be one of the main contributors to squeezing degradation at low frequency, in the region over which the squeezing ellipse rotation takes place, limiting the quantum noise reduction in the radiation pressure dominated region. 
Since they are mainly caused by scattering from mirror defects, the requirements on the mirror quality have been carefully set after performing a complete squeezing degradation budget \cite{mio}. We fixed a requirement of 80 ppm for the round trip losses. This level, combined with other sources of squeezing degradation, whose expected levels are reported in Tab.\ref{valqui}, should allow for a squeezing level of about 4 dB at low frequency and 6 dB at high frequency \cite{mio}. 
The threshold of 80 ppm has been chosen because with this value the associated squeezing degradation at low frequency is comparable with the one expected from the optimistic level of mode mismatching, reported in Tab.\ref{valqui}, therefore lower losses would not increase the squeezing level, unless the mismatching is also reduced.\\
 \begin{table}[!ht]
 \begin{center}    
 \begin{ruledtabular}
 \begin{tabular}{llr}

Squeezing degradation parameter                                                                                                         & Value \\
\hline

Filter cavity losses                                                                                            & $80$ ppm   \\
Injection losses                                                                                               & $5\% $   \\
Readout losses                                                                                              & $5\% $            \\
Mode-mismatch squeezer-filter cavity                                                           & $2\% $    \\
Mode-mismatch squeezer-local oscillator                                                     & $5\% $   \\
Filter cavity length noise (RMS)                                                                                                    &$0.3$ pm \\
Injected squeezing                                                                                  & $9\,\mathrm{dB}  $  \\

\end{tabular}
\end{ruledtabular}
\caption{Parameters used in the estimation of squeezing degradation done in \cite{mio} allowing to reach 4dB of squeezing at low frequency.}
\label{valqui}
\end{center}
\end{table}  

\begin{figure}[htb]
\begin{center}
\includegraphics [scale = 0.29] {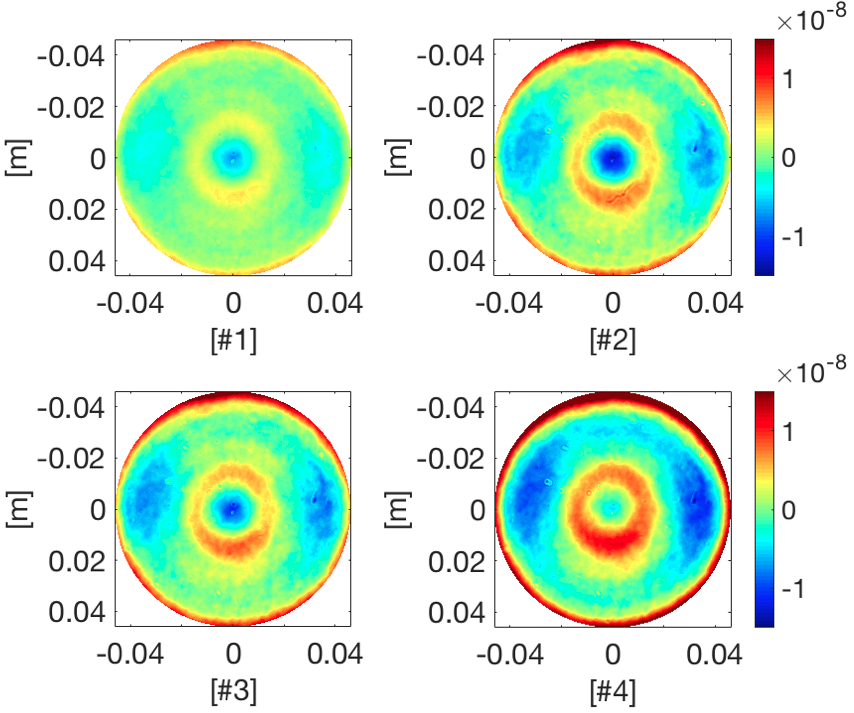}
\caption{Measured flatness maps of the four filter cavity mirrors,. Mirror $\#1$ and $\#4$  have been installed.}
\label{realmapLMA}
\end{center}
\end{figure}

\begin{figure}[h!]
	\centering{\includegraphics[width=0.5\textwidth]{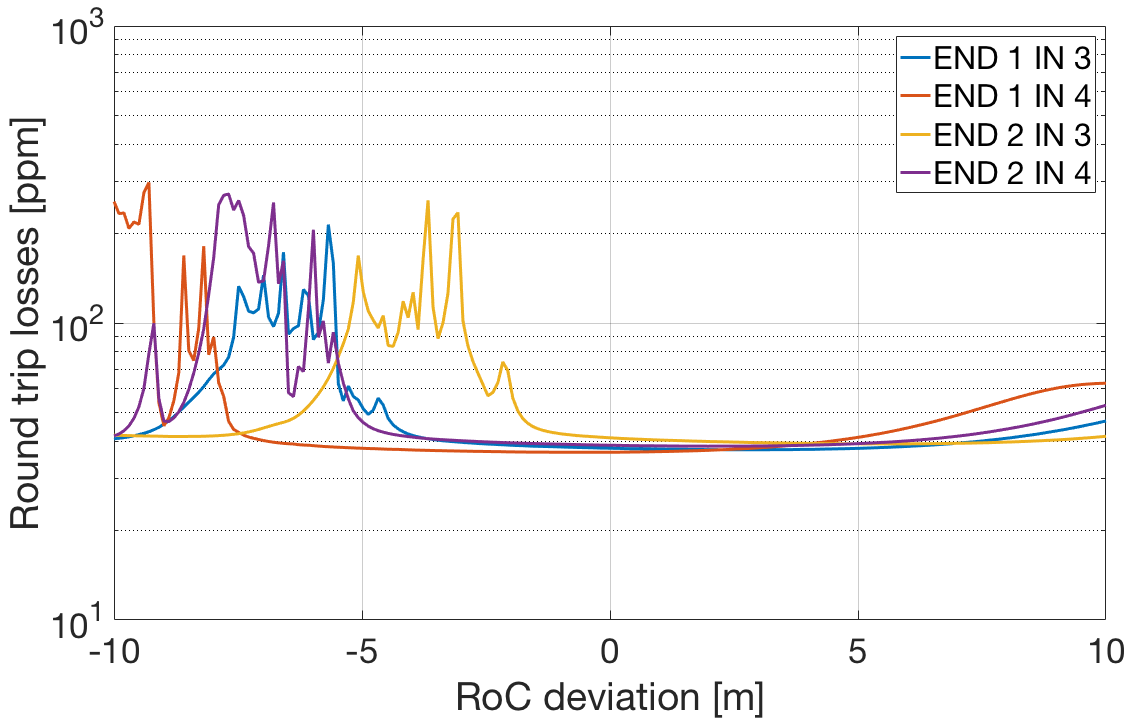}}
	\caption{Round trip losses for different combinations of filter cavity mirrors as a function of the deviation from the measured RoCs. The peaks are due to higher-order modes partially resonant for some values of the two curvature radii. The combination of mirrors $\#1$ and $\#4$ is thus optimal from this point of view.}
	\label{RTLfin}
\end{figure}

The value of 80 ppm includes losses induced by low spatial frequency defects, up to $10^3 \no\,{m}^{-1}$, (contributing to the so-called mirror \textit{flatness}), the losses due to higher frequency defects (contributing to the so-called mirror \textit{roughness}) and the ones due to point defects. The first ones are estimated by performing simulations with maps usually obtained from wave-front measurements with a phase-shifting interferometer \cite{nicolas}, while the latter can be directly measured by recording the scattered light at angles larger than few degrees.\\
By performing Fast Fourier Transform (FFT) simulations with real mirror maps, using the Matlab based optical FFT code OSCAR \cite{oscar}, we set the specification on the mirror peak-valley (PV) to be less than 12.7 nm on a diameter of 0.05 m and less than 6.3 nm on a diameter of 0.02 m. \\ Four mirrors have been purchased for the filter cavity and they have been coated and characterized by LMA in Lyon. The measured maps are plotted in Fig. \ref{realmapLMA}. The results of this characterization are reported in Tab.\ref{recapmes} and show that the mirror flatness is compliant with our requirements. \\
The pair of mirrors to be installed has been chosen performing FFT simulations of the cavity using the measured mirror maps. The round trip losses for the four combinations are plotted in Fig. \ref{RTLfin} as a function of the deviation from the nominal radius of curvature. They all show a round trip losses floor of $\sim 40$ ppm which is compliant with our requirement.  They also show peaks in the losses due to power transferred to higher-order modes that are partially resonant for some values of the two curvature radii.
We have chosen the pair for which the peaks in the losses due to higher-orders mode resonances were more distant from the nominal RoC value. It corresponds to input mirror number 4 and end mirror number 1 (red line in Fig.\ref{RTLfin}).\\
To have a complete estimation of the round trip losses, we should add up all the losses: The flatness given from the simulation ($\sim$40 ppm), measured roughness and point defects ($\sim$5 ppm and $\sim$9 ppm per mirror, respectively), and transmission and absorption from the end mirror ($\sim$5 ppm).
Therefore, the total round trip losses are expected to be  $\sim$ 60 ppm.

\begin{table}[hbtp]
\scalebox{1} {
\begin{tabular}{|l|l|l|l|l|l|l|l|l|l|}
\hline
     &  \multicolumn{2}{c|}{\begin{tabular}[c]{@{}c@{}}diameter\\ 0.05 m\end{tabular}}                                              & \multicolumn{2}{c|}{\begin{tabular}[c]{@{}c@{}}diameter\\ 0.02 m\end{tabular}}                                                                                                          \\ \hline
Mirror & \begin{tabular}[c]{@{}c@{}}RMS\\ (nm)\end{tabular} & \begin{tabular}[c]{@{}c@{}}PV \\ (nm)\end{tabular} & \multicolumn{1}{l|}{\begin{tabular}[c]{@{}l@{}}RMS\\ (nm)\end{tabular}} & \begin{tabular}[c]{@{}c@{}}PV\\ (nm)\end{tabular} \\ \hline
$\#1$     & 2.0                                           & 11.5                                             & 0.52                                                                   & 3.3                                                                                                                                                                           \\ \hline
$\#2$     & 2.1                                          & 12.2                                              & 0.52                                                                  & 3.3                                                                                                                                                                  \\ \hline
$\#3$     & 1.5                                          & 8.3                                             & 0.48                                                                   & 3.4                                                                                                                                                                      \\ \hline
$\#4$     & 1.9                                            & 14.8                                              & 0.48                                                                  & 3.4                                                                                                                                                                     \\ \hline
\end{tabular}}
\caption{RMS and  PV (over different diameters) for the four mirror substrates measured at LMA.}
\label{recapmes}
\end{table}

\section{\label{sec:reflected} Round Trip Losses measurements }    
\begin{figure}[htbp]
\begin{center}
\includegraphics[scale=0.21] {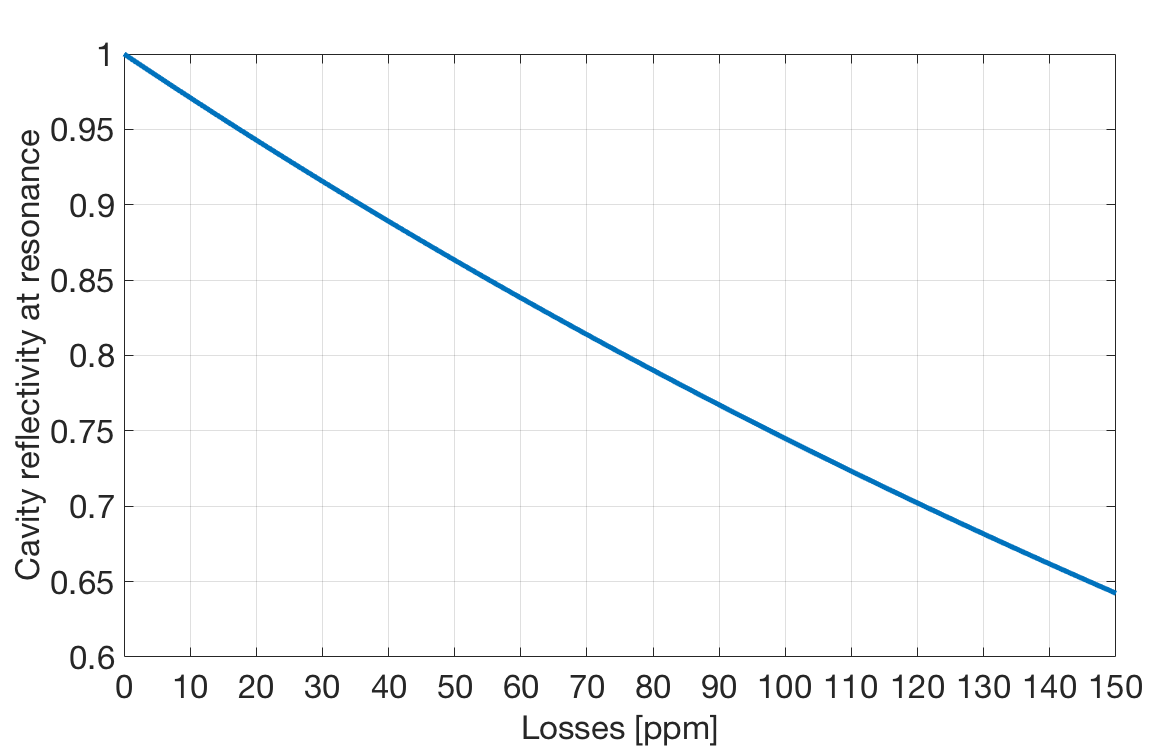}
\caption{Change of the cavity reflectivity as a function of the round trip losses.}
\label{pref_los}
\end{center}
\end{figure}

The round trip losses in a Fabry-Perot cavity are defined via the energy conservation as \cite{rtl_ref}:
\begin {equation} \label{RTLi}
\Lambda^{2}_{rt} = \frac{P_{in}-P_{r}-P_{t}}{P_{circ}}
\end{equation} 
where $P_{in}$ is the input power, $P_{circ}$, $P_{t}$, and $P_{r}$ are the powers circulating in the cavity, transmitted and reflected, respectively.
They affect different optical parameters such as the finesse, the decay time and the cavity power reflectivity which are defined respectively as \cite{isogai}:
\begin{align}\label{f_dct_refl}
F &= \frac{\pi\sqrt{r_1r_2}}{1-r_1r_2}\\
\tau &= -\frac{1}{\nu_{\rm{FSR}}\, \log(r_1r_2)}\\
R &= \left[\frac{r_1-r_2}{1-r_1r_2}\right]^2
\end{align}where $ r_1$ and $r_2$ are the amplitude reflectivities of the input and end mirrors and $\nu_{\rm{FSR}} $ is the free spectral range of the cavity defined as $c/2l$ (with $c$ the speed of light and $l$ the length of the cavity).\\
In principle, the losses can be extrapolated from any of these quantities, for example by incorporating them in the end mirror transmissivity (which in our case is below 4 ppm) \cite{laval}:
\begin{equation}\label{L_T}
r_2 = \sqrt{1-T_2}\rightarrow \sqrt{1-L}
\end{equation}
where $L$ contains also the losses due to the transmission of the end mirror. 
In practice, the extraction of the losses using the finesse and the decay time is limited by the effect of the uncertainty on the input mirror transmissivity. In fact, a relative uncertainty of $1\%$ on the input mirror transmission (corresponding to $\Delta T \sim 0.0014 \%$) results in an error of $\pm15$ ppm of losses.
Luckily, in a strongly overcoupled cavity, as in our case, the cavity reflection is the suitable quantity to measure to measure the round trip losses independently as it has only small dependence on the input mirror transmission.\\ 
The reflectivity of a the cavity at resonance, with the approximation of Eq. \ref{L_T}, reads:
\begin{equation}\label{presn}
R_{\m{cav}} = \frac{P_{\m{res}} }{P_{\m{in}} } = \left[\frac{r_1-r_2}{1-r_1r_2}\right]^2 \simeq \left[\frac{r_1-\sqrt{1-L}}{1-r_1\sqrt{1-L}}\right]^2  
\end{equation}
where $P_{\m{res}}$ is the reflected power on resonance and $P_{\m{in}}$ is the incident power, which is estimated by measuring the reflected power while the cavity is set to be out of resonance.\\
The expected change in the cavity reflectivity induced by losses in our filter cavity is plotted in Fig. \ref{pref_los}. We see that, for example, a change of $3\%$ in the reflectivity corresponds to 10 ppm of losses.\\
The Eq. \ref{presn} can be inverted and approximated \cite{phd_eleonora} to find: 
\begin{equation}\label{aplos}
L \sim \frac{T_1}{2}\frac{1-R_{\m{cav}} }{1+R_{\m{cav}} }  = \frac{T_1}{2}\frac{1-P_{\m{res}} /P_{\m{in}} }{1+P_{\m{res}} /P_{\m{in}} } 
\end{equation}
The losses are measured by repeatedly setting the IR beam on and off resonance, recording the consequent change in the reflected power, as shown in Fig. \ref{lock_unlock}, and computing the reflectivity as the ratio between the reflected power in the two states.
\subsubsection*{Influence of not perfectly coupled input beam}

If a part of the incoming power does not couple with the cavity, (for example in the presence of mismatching, misalignment, modulation sidebands and residual frequency fluctuations due to the finite gain of the locking servo), it will be promptly reflected and will not experience losses. As a consequence, the apparent reflectivity of the cavity on resonance will increase and the measured losses are reduced.\\
Assuming that the round trip losses associated to a cavity are those experienced by an input beam perfectly coupled into the cavity, we are interested in compensating the effect of not coupled power in our reflectivity measurement. Suppose that a fraction $\gamma$ of the incoming power does not couple with the cavity, the reflected power on resonance can be rewritten as:
\begin{equation}
P^{\gamma}_{\m{res}} = R_{\m{cav}} P_{\m{in}}(1-\gamma) + \gamma P_{\m{in}} \\
\end{equation}
From which we can find an expression for $R^{\gamma}_{\m{cav}}$, the cavity reflectivity in the presence of some not coupled power:
\begin{equation}
R^{\gamma}_{\m{cav}} =\frac{P^{\gamma}_{\m{res}}}{P_{\m{in}}}  = R_{\m{cav}} (1-\gamma) + \gamma 
\end{equation}
Inverting the formula above we find the relation to be used to deduce the "real" cavity reflectivity, i.e the one found with a perfectly coupled input beam, knowing the level of not coupled power $\gamma$ and the measured reflectivity $R^{\gamma}_{\m{cav}}$:
\begin{equation}\label{wgamma}
R_{\m{cav}}  = \frac{R^{\gamma}_{\m{cav}}-\gamma}{(1-\gamma) } 
\end{equation}

\begin{figure}[htbp]
\begin{center}
\includegraphics[scale=0.21] {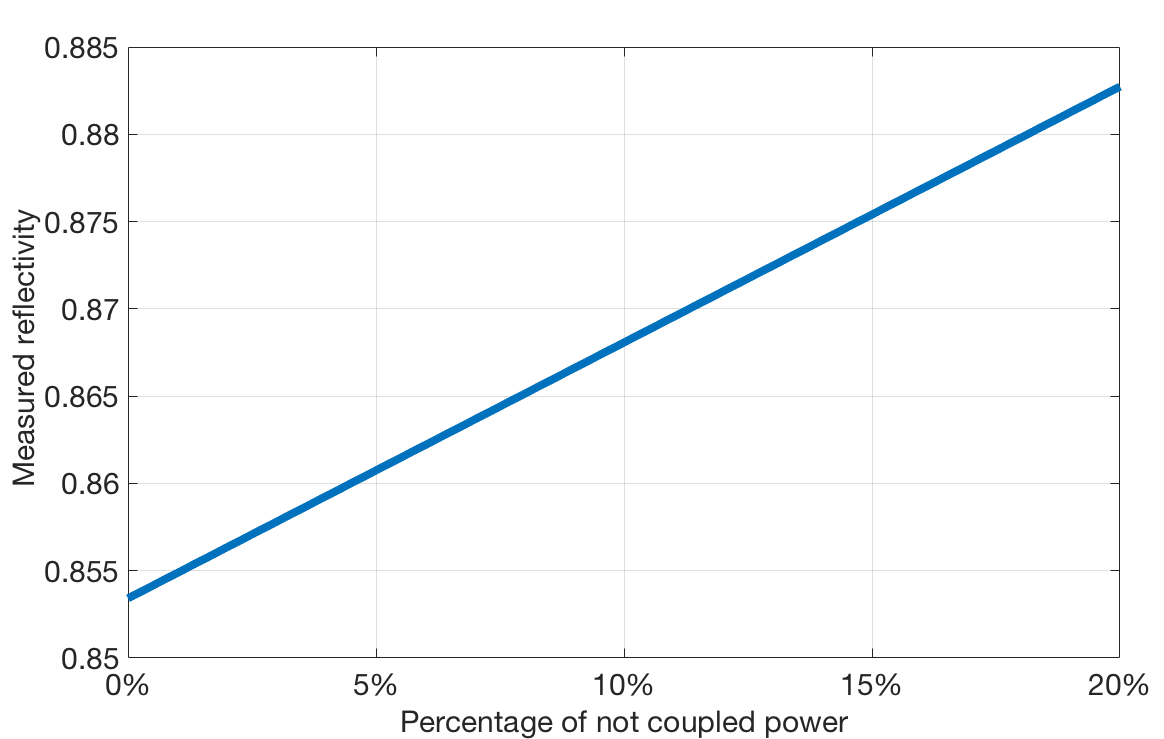}
\caption{Change of the measured cavity reflectivity as a function of the fraction of input power not coupled into the cavity, assuming 60 ppm of round trip losses. }
\label{mmpw}
\end{center}
\end{figure}

The change of the measured cavity reflectivity as a function of the percentage of input power not coupled into the cavity is shown in Fig. \ref{mmpw}. The amount of mismatching and misalignment was estimated to be $\sim 4 \%$ by measuring the optical spectrum of the cavity and comparing the height of the higher orders modes with that of the fundamental one. The power on the radio-frequency sidebands used to generate the Pound-Drever-Hall signal is $\sim 8 \%$ and the power lost for the residual fluctuations of the laser frequency is $\sim 1\%$. All these contributions give  $\sim 13 \%$ of not coupled power in the cavity, whose effect has been compensated in the reflectivity computation with the technique explained above.

\subsubsection*{Data Analysis and results}

\begin{figure}[htbp]
\begin{center}
\includegraphics[scale=0.21] {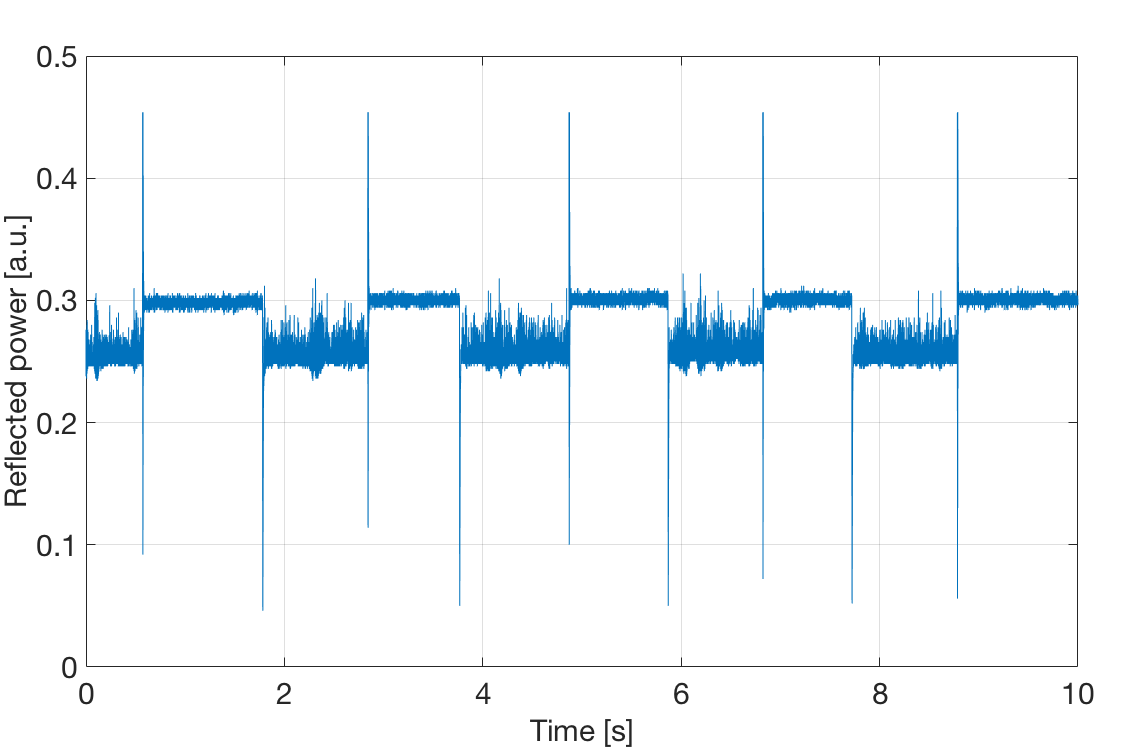}
\caption{Reflected power change during a set of on/off resonance switches of the IR beam. From the difference between the two levels we can estimate the cavity reflectivity and therefore the round trip losses.}
\label{lock_unlock}
\end{center}
\end{figure}

The off/on resonances measurements have been performed several times keeping the cavity locked with the green beam and applying with the AOM a frequency shift large enough to bring the IR beam out of resonance. Fig. \ref{lock_unlock} shows an example of the change in the reflected power when performing such a measurement. For each on/off resonance switch, the cavity reflectivity, and then the losses, has been estimated from the measured ratio between the reflected powers in the resonant and not-resonant state. 

\begin{figure}[htbp]
\begin{center}
\includegraphics[scale=0.21] {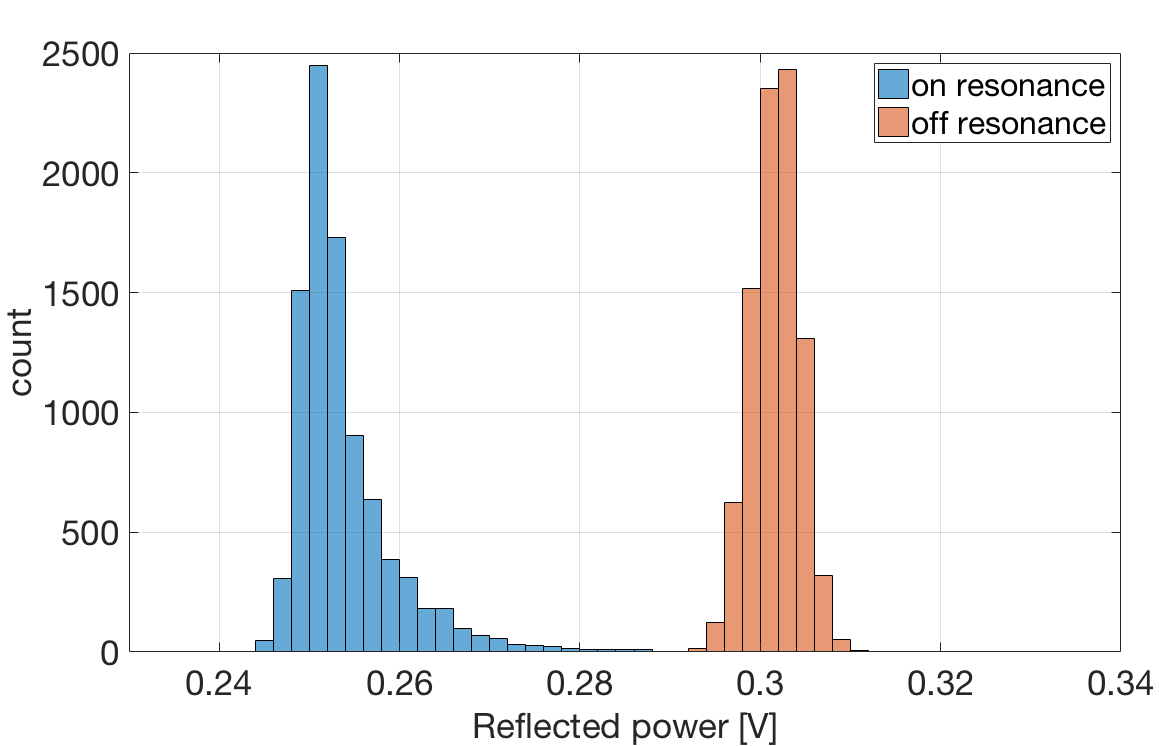}
\caption{A typical histogram of the reflected power time series both for a resonant and for a not resonant period. The fluctuations when the  beam is not resonant are gaussian while those when it is resonant show a broader asymmetric distribution.}
\label{Gauss}
\end{center}
\end{figure}

The reflected power in when the beam is not resonant shows Gaussian fluctuations, as shown in Fig. \ref{Gauss}, mainly due to input power fluctuations, and its level can be estimated taking the mean of the time series, with two standard deviations as uncertainty. On the other hand, the reflected power during the resonant status is subjected to additional fluctuations due to fast alignment  fluctuations, to finite locking accuracy, and possibly to other unknown sources. 

Concerning the alignment fluctuations, if the cavity is not set on the perfect alignment condition, the alignment fluctuations can either increase or reduce the reflected power. On the contrary, since we assume to be locked on the top of the resonance, the fluctuations of the locking point of the cavity (due to the finite lock accuracy), can only increase the reflected power, and therefore are not expected to give a symmetric distribution of the cavity reflectivity. 

We observe that the fluctuations of the reflected light when the beam is resonant usually show an asymmetric distribution. As shown in Fig. \ref{Gauss} the left tail, with respect to the maximum value (the so-called \textit{mode}) is gaussian, with a standard deviation similar to the one of the not resonate state. The right tail (higher powers) is much broader. 

Since the fluctuations toward the higher power (as those of the lock accuracy) bring to an underestimation of the round trip losses, we decided to keep a conservative approach. The highest value of the histogram is used as representative value of the reflected power. The uncertainty of the measurement is chosen as two standard deviations of the gaussian distribution computed using only the left tail (assuming that the left tail given by power and alignment gaussian fluctuations). It is important to point out that using the mode instead of the mean, gives a difference of only $\sim$ 5 ppm in the estimated losses, meaning that our analysis is not strongly dependent on this choice.

\begin{figure}[htbp]
\begin{center}
\includegraphics[scale=0.215] {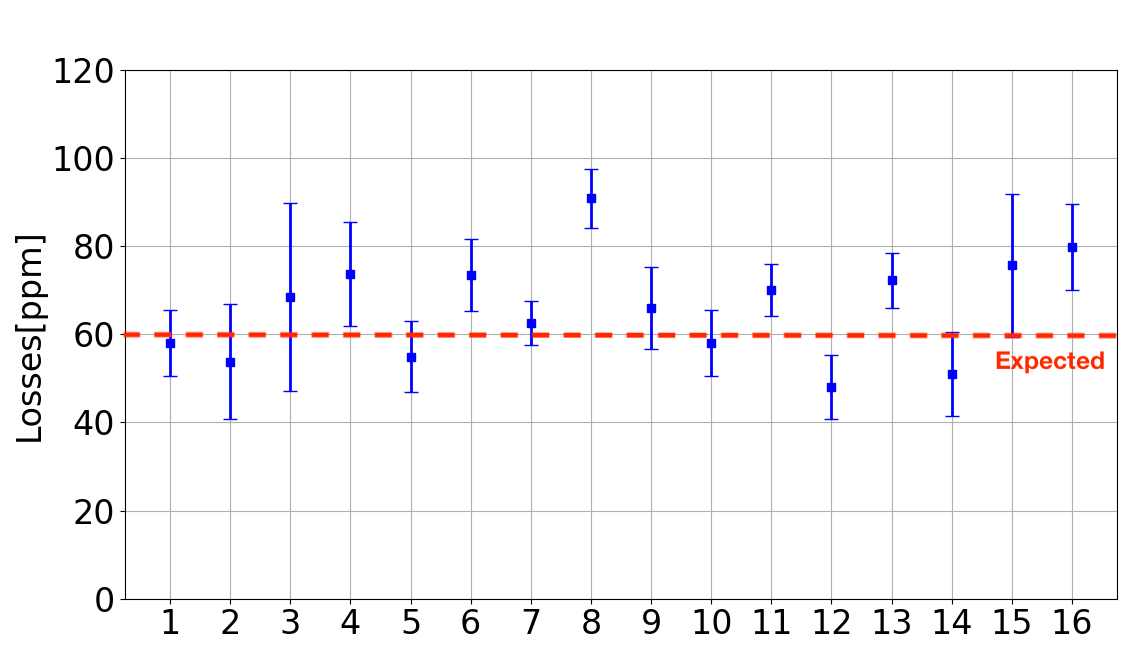}
\caption{Summary of the round trip losses measurements. The scattering of the results may depend on different alignment conditions of the cavity from one day to another. The measurements have been taken on different days during about two months.}
\label{sum_res}
\end{center}
\end{figure}

For each measurement (i.e. a set of consecutive on/off resonance switches, as the one in Fig. \ref{lock_unlock} ) we have computed the reflectivity as the mean (weighted with uncertainties) of the reflectivities obtained from each on/off resonance switch, with the method described above. The results are plotted together in Fig. \ref{sum_res}.

\begin{figure}[h!]
\begin{center}
\includegraphics[scale = 0.27]{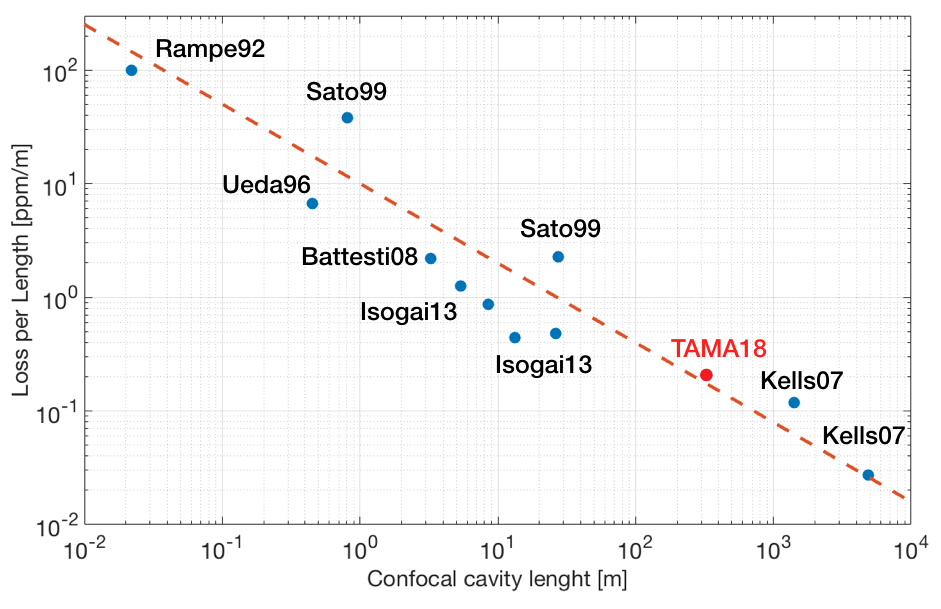}
\caption{The plot shows some measured round-trip loss per unit length from the literature. It was originally published in \cite{evansfc}, than updated with measurements by Isogai et al. in \cite{isogai}. We added the measured losses for filter cavity in TAMA (this work). To remove any dependence on the choice of cavity geometry the plots are done in function of the confocal length, i.e  the length of the confocal cavity which has the same beam dimension on the mirror as the cavity whose losses are reported. References for the measurements in literature can be found in \cite{evansfc}.}
\label{isogaiplot}
\end{center}
\end{figure}

The measured losses are between 50 and 90 ppm (with typical error bars between $\pm 5$  and $\pm 10$ ppm) while from simulation we expected $\sim$ 60 ppm. The fluctuation of the reflectivity from one measurement to another is larger than the experimental uncertainty of each of them. This phenomenon has been already observed in \cite{isogai} and it is possibly due to different alignment conditions: the beam impinges on different areas of the mirrors that can have slightly different surface quality causing a variation in the amount of scattered light. The mean of these measurements gives $ \sim 67$ ppm of losses (0.22 ppm/m). We plotted this value together with other measured RTL per unit length from the literature in Fig. \ref{isogaiplot}. It shows a good agreement with the empirical scaling law (dotted line in Fig. \ref{isogaiplot}) extrapolated from the previous measurements.\cite{isogai}.
%\subsection{\label{sec:reflected} Measurements using cavity's reflected power}    

%\subsection{\label{sec:decay} Measurements using cavity's decay time} 

\section{\label{sec:conclusion} Conclusions and next steps} 

\begin{figure}[h!]
\begin{center}
\includegraphics[scale = 0.15]{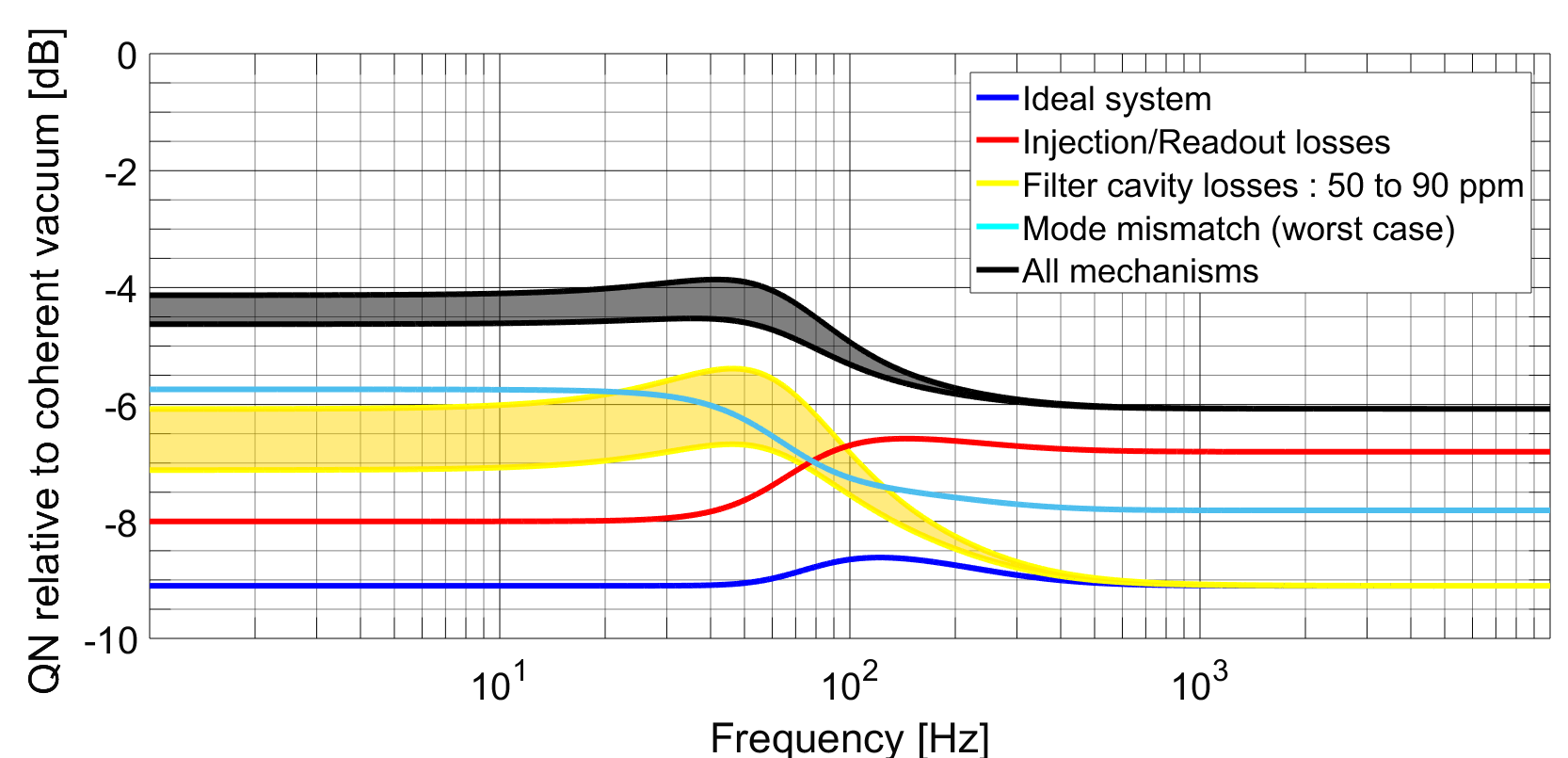}
\caption{Quantum noise relative to coherent vacuum (0 dB), computed taking into account different squeezing degradation mechanisms. The yellow area corresponds to the contribution from filter cavity losses between 50 and 90 ppm. The parameters used to estimate the other degradation mechanisms are those reported in Tab. \ref{valqui}}
\label{sqz_deg}
\end{center}
\end{figure}

We have shown the operation of a 300 m filter cavity designed to impress a quadrature rotation at about 70 Hz of the vacuum squeezed state, that is to be injected in future upgrades of LIGO, Virgo, and KAGRA. The cavity has been controlled by using a green beam, obtained by doubling the frequency of the infrared laser used for the squeezing generation. We have been able to control the resonance condition of an infrared probe beam with an AOM. In this configuration, we have been able to estimate the round trip losses of the cavity, which are the most critical parameter affecting the squeezing level in the region radiation-pressure dominated.  From the measurement of the cavity reflectivity, with the on/off resonance technique, we obtained round trip losses between 50 and 90 ppm, where the scattering of the results may depend on different alignment conditions of the cavity. The squeezing degree achievable with this losses level is plotted in Fig. \ref{sqz_deg}, where the parameters used to estimate the other degradation mechanisms are those reported in Tab. \ref{valqui}. Even with the worst results, the losses are compatible with 4 dB of squeezing in the region radiation-pressure dominated and 6 dB of squeezing at high frequency. \\
The installation of the frequency independent squeezing source is ongoing as well as the integration of an automatic alignment system, which we expect to improve the stability of the cavity and the precision and reproducibility of the losses measurement. 

\FloatBarrier

\section*{Acknowledgements}
We thank J\'er\^{o}me Degallaix for fruitful discussions about the loss measurement and the help with OSCAR simulations.  We thank also Advanced Technology Center of NAOJ for the support. This work was supported by the JSPS Grant-in-Aid for Scientific Research (Grant code 15H02095), the JSPS Core-to-Core Program, A. Advanced Research Networks, and the European Commission under the Framework Program 7 (FP7) 'People' project ELiTES (Grant Agreement No. 295153) and EU Horizon 2020 Research and Innovation Programme under the Marie Sklodowska-Curie Grant Agreement No. 734303. E. C. was supported  by the European Gravitational Observatory, by the scholarship "For Women in Science" from the Fondation l'Or\'eal UNESCO and by the scholarship "Walter Zellidja" from the Acad\'emie Fran\c{c}aise.
%\bibliographystyle{utphys}

%\bibliography{biblio}

\end{document}